# Electronic Coupling between the Unoccupied States of the Organic and Inorganic Sub-Lattices of Methylammonium Lead Iodide a Hybrid Organic-Inorganic Perovskite Single Crystal

Gabriel J. Man[1]*, Cody M. Sterling[2], Chinnathambi Kamal[2,3], Konstantin A. Simonov[1,+], Sebastian Svanström[1], Joydev Acharya[4], Fredrik O.L. Johansson[1], Erika Giangrisostomi[5], Ruslan Ovsyannikov[5], Thomas Huthwelker[6], Sergei M. Butorin[1], Pabitra K. Nayak[4]*, Michael Odelius[2], Håkan Rensmo[1]*

**Author addresses:**

1. Condensed Matter Physics of Energy Materials, Division of X-ray Photon Science, Department of Physics and Astronomy, Uppsala University, Box 516, Uppsala, 75121, Sweden

2. Department of Physics, Stockholm University, AlbaNova University Center, Stockholm, 10691, Sweden

3. Theory and Simulations Laboratory, HRDS, Raja Ramanna Centre for Advanced Technology, Indore, 452013, India

4. TIFR Centre for Interdisciplinary Sciences, Tata Institute of Fundamental Research, 36/P, Gopanpally Village, Serilingampally Mandal, Hyderabad, 500046, India

5. Institute Methods and Instrumentation for Synchrotron Radiation Research, Helmholtz-Zentrum Berlin GmbH, Albert-Einstein-Strafze 15, Berlin, 12489, Germany

6. Paul Scherrer Institut, WLGA/212, Forschungsstrasse 111, 5232 Villigen, Switzerland

+ Present address: Swerim AB, Kista, Stockholm, 16407, Sweden

* To whom correspondence should be addressed.  Email:

gman@alumni.princeton.edu

pabitra.nayak@tifrh.res.in

hakan.rensmo@physics.uu.se

**Abstract**

Organic-inorganic halide perovskites have been intensively re-investigated due to their applications, yet the opto-electronic function of the organic cation remains unclear.  Through organic-selective resonant Auger electron spectroscopy measurements on well-defined single crystal surfaces, we find evidence for electronic coupling in the unoccupied states between the organic and inorganic sub-lattices of the prototypical hybrid perovskite, which is contrary to the notion based on previous studies that the organic cation is electronically inert.  The coupling is relevant for electron dynamics in the material and for understanding opto-electronic functionality.

In spite of the substantial amount of research effort invested globally, the opto-electronic function of the organic cation in the prototypical hybrid organic-inorganic perovskite (HOIP) methylammonium lead tri-iodide ($MAPbI_3$ or MAPI) is still unclear [1,2]. Since the early demonstrations of methylammonium-containing perovskite solar cells, device applications of halide perovskites (HaPs) have transitioned towards the use of mixed-cation (with/without methylammonium) formulations due to concerns about the thermal and chemical stabilities of methylammonium [3,4]. This trend was justified in part by evidence from a multitude of approaches that the organic cation is predominantly an electronically-inactive filler [2,5–12]. However, the highest-performing HaP solar cells typically incorporated some fraction of methylammonium in the absorber layer and hot fluorescence was only observed from lead bromide perovskites with organic cations, suggesting that the organic sub-lattice plays an opto-electronic role [13–15]. In principle, the comparison between direct or inverse photoelectron spectroscopy measurements and electronic structure calculations should reveal the contributions from the occupied or unoccupied states of the organic and inorganic sub-lattices [6,16]. The occupied states near the valence band maximum (VBM), relevant for opto-electronic functionality in MAPI and related compounds, have been studied using Angle-resolved Photoelectron Spectroscopy (ARPES); however, it is unclear if methylammonium-related states are present in the energetic region close to the VBM as no explicit evidence for their presence has been found using non-element-selective PES [17,18]. The unoccupied states have been less commonly investigated; a combination of non-element-selective Inverse Photoelectron Spectroscopy (IPES) and density-of-states (DOS) calculations suggests the unoccupied methylammonium states are energetically positioned near the vacuum level and may not play a role in opto-electronic functionality involving band-edge states [6,19].

In this Letter, we investigate both the valence band and conduction band states of MAPI via an array of electron spectroscopies performed on clean surfaces of in-vacuum cleaved single crystals, and find evidence for organic-inorganic electronic coupling in the unoccupied states. We have resonantly excited the nitrogen *1s* (N *1s*) core level to impart organic cation-selectivity to the electron spectroscopies. We have performed checks for beam-induced sample damage and ensured the measurements are representative of the intrinsic physics of MAPI. Our nitrogen-projected DOS calculation suggests a spectroscopically low intensity of methylammonium states, relative to the iodide states, exists in the valence band region. Our N *1s* Resonant Photoelectron Spectroscopy (RPES) measurements confirm that the intensity of methylammonium states, if present in the valence band, is low, and are consistent with a report which shows a low intensity of methylammonium states near the VBM using a complementary technique (N *1s* X-ray Emission Spectroscopy (XES)) [20]. The low intensity distribution of methylammonium states may still be relevant for opto-electronic functionality. In the course of performing N *1s* RPES, we observe nitrogen core-valence-valence (N *KVV*) Auger decay, contrary to the findings of a study which reports N *1s* RPES measurements on MAPI films [21]. Through the use of N *1s* Resonant Auger Electron Spectroscopy (RAES), we discover the existence of X-ray-excited electron de-localization with a timescale comparable to the N *1s* core hole lifetime of ~6 fs, demonstrating electronic coupling between the unoccupied states of the organic and inorganic sub-lattices. Based on the aforementioned findings, we conclude that the occupied states of the organic and inorganic sub-lattices of MAPI are largely independent but the unoccupied states are significantly coupled, and suggest that the coupling is relevant for slow hot electron cooling.

Beam-induced chemical changes to HaP sample surfaces is a known experimental challenge; the starting composition of a HaP film has been reported to influence how tolerant HaP samples are to X-ray irradiation [22,23]. We mitigate this experimental challenge through the use of single crystals, which offer a lower concentration of processing-related defects, and a low photon intensity beamline (~$10^{11}$ photons cm$^{-2}$ s$^{-1}$), which necessitates the use of a high transmission electron spectrometer [24]. Millimeter-scale single crystals of MAPI were grown in solution and characterized with X-ray diffraction (XRD), which confirmed the presence of the tetragonal phase at room temperature and its purity [25,26]. Electron spectroscopy measurements were performed at the synchrotron BESSY II, at the soft X-ray beamline PM4 and end-station LowDosePES [24]. The crystals were cleaved in ultra-high vacuum (UHV) immediately prior to measurement. All measurements were performed at room temperature in a UHV chamber with a base pressure of 1x10$^{-9}$ mbar. Using PES measurements, we have extensively checked that the chemical integrity of the single crystal surfaces remains intact [26]. Through the use of clean surfaces, we ensured all electron emission arising from N *1s* core hole decay can be attributed solely to methylammonium; the N *1s* PES spectrum can be fitted with one symmetric peak which we assign to the nitrogen atom in methylammonium. Nitrogen *K*-edge Near Edge X-ray Absorption Fine Structure (NEXAFS) and ground-state electronic structure calculations were performed on snapshots of the 2x2x2 supercell geometry sampled for five configurations at regular intervals during the Ab Initio Molecular Dynamics (AIMD) simulations, where the initial supercell model was generated from reported lattice parameters [25]. The AIMD simulations were performed with the CP2K code at 300 K and 0 atmosphere (atm) to approximate the experimental conditions [27]. Nitrogen *K*-edge NEXAFS spectra, obtained within the half core-hole transition potential approximation (TPHH), were calculated for each nitrogen present in the supercell [28]. Projected DOS calculations were performed using the ground-state wavefunction. The reader is referred to the Supplemental Material for basic characterization (X-ray diffraction, surface chemical integrity checks and characterization with PES) and further experimental and computational details [26].

The complex composition of MAPI implies that element- and orbital-projected DOS calculations are essential for interpreting valence band photoemission spectra, and the agreement depends on the underlying structural model, the photo-ionization cross-sections and the modeling of the spectroscopic process [26]. We compare experimental and computed values of the VB-peak-to-NEXAFS-peak energy offset, shown in Fig. 1, and simultaneously examine the accuracies of the underlying structural model, the simulated NEXAFS spectrum and the modeling of the PES spectra in terms of DOS calculations. The experimental spectra consist of the valence band PES and N *K*-edge NEXAFS spectra. The NEXAFS spectrum was aligned onto the PES energy scale by subtracting the photon energy from the N *1s* binding energy. The valence band spectrum of MAPI is similar to a reported spectrum recorded with hv = 500 eV [16]. The computed spectra include the I *p*-projected DOS as a representation of the occupied states, since it comprises the dominant contribution to the valence band, and the computed N *K*-edge NEXAFS (aligned using the computed N *1s* orbital energy in the TPHH calculation). The ground-state DOS calculations are consistent with previous reports and serve as an approximation to the photoemission spectra [26]. We observe that the VB-peak-to-NEXAFS-peak offsets for the experimental and theoretical spectra are in good agreement, demonstrating that our structural model and calculations are sufficiently accurate. Changes to the Fermi energy position would affect the VB and NEXAFS (via the N *1s* binding energy) positions equally. The computed N *p*-projected DOS of MAPI is displayed also in Fig. 1 as a first approximation to the N *K*-edge NEXAFS spectrum. Core hole relaxation effects are manifested as a build-up of resonance intensity towards higher binding energy and a narrowing of the DOS, as evidenced by the comparison between the ground-state nitrogen *p*-projected DOS and the experimental and simulated N *K*-edge NEXAFS spectra. Other details displayed in Fig. 1 include the reported band-gap of

MAPI (~1.6 eV) and the energy position of the ionization threshold (or vacuum level), which is estimated from the experimentally determined position of the valence band maximum and reported values for the band-gap (1.6 eV) and electron affinity (3.6 eV) [6,26]. Detailed electron spectra were recorded at the five photon energies labelled in the plot. The deduced energy position of the ionization threshold suggests that photo-excitations with energies ≤ 405.3 eV are to bound states while photo-excitations with energies ≥ 407.0 eV are to continuum states. This deduction is consistent with the observation that the normal Auger spectrum (hv = 412.3 eV) is similar to the hv = 407.0 eV spectrum, within signal-to-noise limitations [26]. Due to its higher signal-to-background, the hv = 407.0 eV spectrum will be utilized as the normal Auger spectrum. We conclude that the accuracy of our computational work is adequate for interpreting experimental valence photoemission spectra. See Supplemental Material for the reference ground-state DOS calculations and supporting data analyses [26].

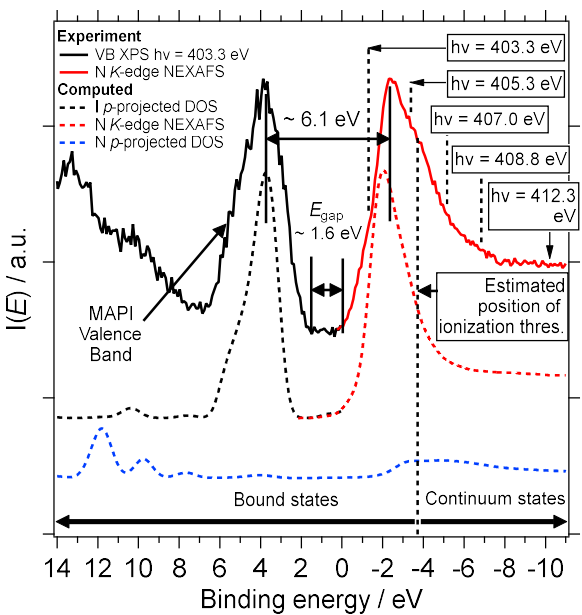

FIG. 1. Experimental versus computed valence photoemission and N $K$-edge NEXAFS spectra aligned onto a common energy scale. See text for details regarding the energy alignment and the estimation of the energetic position of the ionization threshold. The valence band peak to NEXAFS peak offset serves as one common reference between experiment and computation. The photon energy labels show the excitation energies used to record detailed auto-ionization and Auger electron spectra.

In Fig. 2, we present an off-resonant or background spectrum (hv = 394. 5 eV) and a resonant spectrum (hv = 405.3 eV), from which a difference spectrum, which highlights all electron emission channels (with different final states) associated with N $1s$ core hole decay, is extracted. A wider energy region is shown in the Supplemental Material [26]. We compare the N $1s$ participator auto-ionization or resonant photoemission portion, of the difference spectrum, to the computed ground state N-projected DOS. Participator auto-ionization features are identified by their dispersive nature versus varying photon energy; if plotted on a binding energy scale, they remain constant. A comparison of the hv = 405.3 eV (Fig. 2) and 403.3 eV (Fig. S9(c)) difference spectra indicates that similar features in the same 16 - 0 eV binding energy region can be seen. To align the N-projected DOS to the valence band spectra, the I-projected DOS is first aligned in a manner that is consistent with a past report [16]. The nitrogen-projected DOS shows a number of features labeled according to the molecular point group of the methylammonium cation. The highest-occupied-molecular-orbital (HOMO) of methylammonium is the 2e molecular orbital (MO). The calculations suggest that hydrogen-bonding interaction(s) between methylammonium and its environment results in the formation of states with weak nitrogen character at binding energies close to the VBM. We interpret the origin of these states as hybridization between the organic and inorganic sub-lattices and note that the intensity of the hybridized states is low relative to the intensities of the other N features. The hybridized feature has been observed with N *K*-edge XES, a complementary technique which interrogates the occupied states [20]. The participator auto-ionization features have a positive slope as a function of increasing binding energy, which qualitatively agrees with the N-projected DOS. Our RPES measurement reveals that the spectroscopic intensity of the hybridized states, if present, is low. Kot *et al.* performed N *1s* RPES measurements on thin films of MAPI and observed nitrogen-related resonant intensity enhancements close to the VBM, though it is unclear what the origin of the nitrogen states is due to the multiple nitrogen species seen in their N *1s* PES spectrum [21]. The overlap of nitrogen and iodide states near the VBM is small, according to our DOS calculations, suggesting that the occupied electronic sub-structures of the organic and inorganic sub-lattices are largely independent.

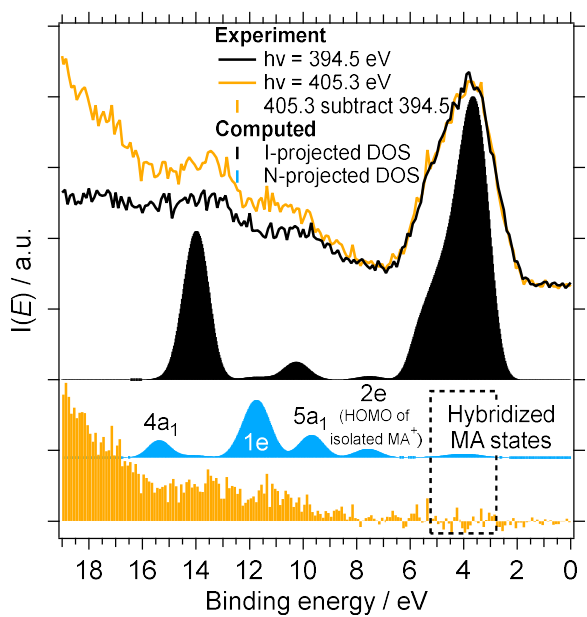

FIG. 2. Nitrogen contributions from the organic sub-lattice to the electron spectra. The computed DOS is aligned to the valence band spectra using the I-projected DOS. The subtraction of the off-resonant hv = 394.5 eV spectrum from the resonant hv = 405.3 eV spectrum yields a difference spectrum for comparison to the nitrogen-projected DOS.

The participator auto-ionization shown in Fig. 2 comprises a minority fraction of the total N *1s* core hole decay.  A comparison of the decay spectra arising from excitations to bound (hv = 403.3 eV) versus continuum (hv = 407.0 eV) states is shown in Fig. 3; it reveals that the majority of the N *1s* decay is similar (normal Auger decay).  This indicates that for the majority of excitations to bound states, the photo-excited electrons have de-localized from their host nitrogen atom and are unable to participate or spectate in the local decay of the N *1s* core hole.  Possible origins of de-localization include resonant excitation directly into a hybridized and extended (Bloch-like) organic-inorganic state and/or transfer of the excited electron from the organic-to-inorganic sub-lattices with a timescale that is clocked by the N *1s* core hole lifetime (~6.3 fs) [29,30].  Both cases imply that the organic and inorganic sub-lattices are electronically coupled.  We interpret the electron de-localization as charge transfer (CT) from the organic to inorganic sub-lattices of MAPI for the purpose of estimating a CT time  [31,32].  We note that the spectral fingerprints of dynamics in the N-H bond, a connector between the sub-lattices, in the intermediate core-ionized state were reported in a N *K* XES study of $MAPbI_{3-x}Cl_x$, which suggests similar dynamics could be present in our RAES measurements involving the core-excited state and affect CT [20].  The relevant decay spectra to examine are the 403.3 eV and 405.3 eV spectra.  The energetic overlap, seen in Fig. 1, between the experimental N *K*-edge NEXAFS and the bound MAPI empty states (not shown) located between the ionization threshold and the CBM suggests that electron transfer between the organic and inorganic sub-lattices is possible, in the N *1s* core-excited state.  The 403.3 spectrum can be regarded as a superposition of the normal Auger spectrum and auto-ionization features.  To highlight the auto-ionization features in the difference spectrum, we normalize the two spectra utilizing the ~362 to ~370 eV region and the background levels at 402 eV.  We observe that the difference spectrum lacks spectral features in the ~362 to ~370 eV region, which indicates that normal Auger decay, the spectral fingerprint of electron de-localization, is present in the hv = 403.3 eV spectrum.  We emphasize that this observation holds even if the difference spectrum is generated from the un-normalized 403.3 and 407.0 spectra.  This interpretation of organic-to-inorganic sub-lattice CT is based in part on the observation that the computed C-, N- and H-projected DOS overlap in the same energy regions (Fig. S8), which implies that the methylammonium orbitals are fully hybridized and therefore CT from one methylammonium MO to another MO seems unlikely.  To estimate the CT time, we assume electron de-localization is a tunneling process with exponential probability as a function of time, the core hole decay rate is exponential also, and the two processes are independent  [30].  The CT time $\tau_{CT} = \tau_{N\ 1s} \times \frac{I_{auto-ionization}}{I_{normal\ Auger}}$ , where $\tau_{N\ 1s}$ is the ~6.3 fs core hole lifetime of N *1s* and *I* represents the integrated area of auto-ionization or normal Auger decay, extracted from the hv = 403.3 eV spectrum.  The normal Auger energy window for ammonium extends over ~50 eV and the Auger energy window for methylammonium is expected to be similarly wide  [33,34].  Given the limited energy windows of the spectra we have recorded, which introduces uncertainties into the modeling of the inelastic background, we provide an order-of-magnitude estimate of the CT time here.  In the range between 362 to 404 eV KE, the integrated area of the normalized hv = 403.3 eV spectrum is 16.2 and the integrated area of the normalized hv = 407.0 eV spectrum is 14.0, with the auto-ionization features accounting for the difference.   The normal Auger component is certainly over-estimated since the inelastic background has been included; hence, we do not provide uncertainty estimates. Our $\tau_{CT} = 6.3\ fs\ \times \frac{2.2}{14} = 1\ fs$ estimate establishes a lower bound for this particular excitation (hv = 403.3 eV). A similar treatment of the 405.3 eV spectrum indicates that the vast majority of the N *1s* decay is Auger and the auto-ionization features are barely visible, leading to a CT estimate that is < 1 fs.  Given our two data points, the CT efficiency appears to be positively correlated with photon energy, and one observation of this finding is that the methylammonium-Pb,I electronic coupling strength increases with more negative binding energies towards the ionization threshold.  The use of RAES for elucidating element-specific features in the conduction band has been demonstrated for other materials.  For single crystal TiO$_2$, crystal field splitting originating from titanium *d*-states in the conduction band was identified [35].  We conclude that normal Auger decay, the spectral fingerprint of electron de-localization, has been found for resonant excitations to bound states, and if interpreted as CT from the organic-to-inorganic sub-lattices of MAPI, leads to an order-of-magnitude estimate of 1 fs for the CT time.  See Supplemental Material for supporting data analyses and a schematic displaying charge transfer, participator and spectator auto-ionization processes [26].

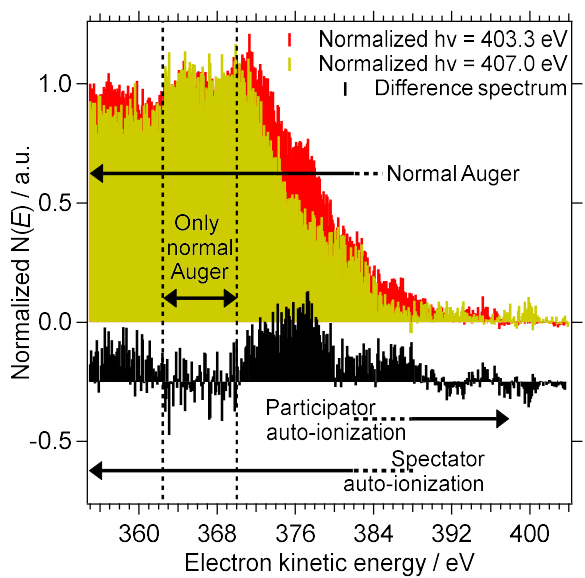

FIG. 3.  Nitrogen *1s* core hole decay spectra associated with resonant excitations to bound (hv = 403.3 eV) versus continuum (hv = 407.0 eV) states.  The hv = 407.0 eV spectrum is treated as a normal Auger spectrum (see text).  The difference spectrum highlights the auto-ionization features.

We first summarize the findings relevant for organic-inorganic electronic coupling in MAPI, then discuss the possible implications for electron dynamics in HOIPs.  We demonstrated chemically well-defined single crystal surfaces of the prototypical HOIP MAPI that are stable under continuous X-ray irradiation and concomitant electron emission for hours.  This development provided access to the intrinsic physics of HOIPs.  We assessed our computational work to be sufficiently accurate, based on a comparison of the experimental and computed VB-peak-to-NEXAFS-peak offsets.  The computed nitrogen-projected DOS shows weak features close to the VBM of MAPI which we attribute to hydrogen bonding between the organic and inorganic sub-lattices.  The computed intensity of the hybridized states is low relative to the other features in the N-projected DOS, which may explain why we have not directly observed the states with N *1s* RPES.

In our investigation of the conduction band states of MAPI, the element specificity of RAES showed, for resonant excitations to bound states (below the core-ionization threshold), that the photo-excited electrons de-localized from their host nitrogen atoms with a timescale comparable to the N *1s* core hole lifetime of 6.3 fs. Irrespective of the origin of de-localization, the presence of de-localization shows the existence of electronic coupling between the conduction band states of the organic and inorganic sub-lattices. By interpreting the de-localization as CT from the X-ray absorption populated states of the organic sub-lattice to the empty states of the inorganic sub-lattice, we deduce the lower bound of CT to be ~1 fs. Charge transfer in the core-excited state, relevant for RAES, may be related to CT in the valence-excited state, relevant for opto-electronic functionality, if the core and valence exciton binding energies are similar. The energies are similar for materials which exhibit efficient core-hole screening such as $C_{60}$ [36].

Our work motivates further inquiry into the mechanism(s) of electron dynamics in HOIPs. The mechanism of slow hot carrier cooling in HaPs is currently not understood and slow hot carrier cooling has been attributed primarily to slow electron cooling [37,38]. Polaronic effects have been associated with the mechanism of slow hot carrier cooling in HaPs; we note that though polaronic effects exist in HaPs, it is not clear if both the polaronic and potential CT effects can operate concurrently [39–41]. The mechanism underpinning slow hot carrier cooling may be related both to the CT mechanism and the structural dynamics of methylammonium, since the femtosecond charge-transfer time is substantially faster than the two characteristic time constants, ~300 fs and ~3 picoseconds, associated with structural re-orientation of the methylammonium cation [42].

**Acknowledgements**

We thank *Helmholtz-Zentrum Berlin für Materialien und Energie* for the allocation of beamtime at PM4 (191-08328 and 192-08712). The research leading to this result has been supported by the project CALIPSOplus under the Grant Agreement 730872 from the EU Framework Programme for Research and Innovation HORIZON 2020.

We acknowledge Svante Svensson (Uppsala), Olle Björneholm (Uppsala), George Sawatzky (UBC) and Reinhold Fink (Tübingen) for fruitful discussions on modeling the N *KVV* AES, and Tomas Edvinsson (Uppsala) for discussions on photo-induced charge transfer.

GJM thanks Lucinda Man for the discussion on energy coupling between cubo-octahedron and the octahedron sub-structures. GJM thanks Sigurd Wagner (Princeton) for the suggestion to investigate the hot carrier cooling mechanisms in lead halide perovskites, and Jeff Schwartz (Princeton) for the discussion on methylammonium de-protonation. GJM thanks Luis K. Ono (OIST) and Yabing Qi (OIST) for their suggestions on how to cleave HaP crystals in-vacuum.

GJM, SS, KAS, HR acknowledge the Swedish Research Council (grant # 2018-06465 and # 2018-04330) and the Swedish Energy Agency (grant # P43549-1) for funding. FOLJ acknowledges the Swedish Research Council (grant # 2014-6463 and # 2018-05336) and Marie Skłodowska Curie Actions (Cofund, Project INCA 600398) for funding.

SMB acknowledges financial support from the Swedish Research Council (grant # 2018-05525).

CS, KC and MO acknowledge support from the European Union's Horizon 2020 research and innovation programme under the Marie Skłodowska-Curie grant agreement No. 860553, and the Swedish Energy Agency (grant # 2017-006797). The calculations were enabled by resources provided by the Swedish National Infrastructure for Computing (SNIC) at the Swedish National Supercomputer Center (NSC), the High Performance Computer Center North (HPC2N), and Chalmers Centre for Computational Science and Engineering (C3SE) partially funded by the Swedish Research Council through grant agreement no. 2018-05973.

PKN acknowledges support from the Department of Atomic Energy, Government of India, under Project Identification no. RTI 4007 and SERB India core research grant (CRG/2020/003877).

**Competing Interests**

The authors declare no conflicts of interest.

SUPPLEMENTAL MATERIAL

## Experimental and Computational details

*MAPI crystal growth.* Single crystals of tetragonal-phase MAPI were prepared via a modification of a reported procedure [43]. Briefly, lead iodide ($PbI_2$, from Sigma-Aldrich) and methylammonium iodide (MAI, from Dyesol) were dissolved into γ-butyrolactone (GBL) to form a solution where the concentration of $PbI_2$ is 0.5 M and that of MAI is 1.5 M. Chloroform (~680 µL/mL) was added to the aforementioned solution. The solution was filtered into a clean, 4 mL glass vial using a 0.45 µm filter. Chloroform was placed into another 4 mL glass vial. Both vials were placed inside a bigger vial, which was then capped and placed into an oil bath kept at 45 °C. This arrangement allowed the chloroform to slowly diffuse into the GBL solution. After two days of incubation, millimeter-scale MAPI crystals had formed. The crystals were collected and preserved in chlorobenzene, and blow-dried with nitrogen prior to use.

*X-ray diffraction measurements.* Measurements were performed on the MAPI crystals at 298 K with a Rigaku diffractometer and graphite-monochromated molybdenum Kα radiation (λ = 0.71073 Å). CrysAlisPro software was used for data processing. Empirical absorption correction was applied to the collected reflections with SCALE3 ABSPACK and integrated with CrysAlisPro. The structure of the MAPI single crystal was solved by direct methods using the SHELXT program, and refined with the full-matrix least-squares method based on $F^2$ by using the SHELXL program through the Olex interface.

*Precursor film preparation.* Thin films of the binary precursor compounds $PbI_2$ and MAI were prepared via spin-coating inside a nitrogen glovebox. High-purity (99.999%) $PbI_2$ was sourced from TCI. The fluorine-doped tin oxide (FTO) coated glass substrates were first ultra-sonicated in acetone, then isopropanol. Residual solvent was removed with dry nitrogen. Solutions of MAI (35 mg/mL in isopropanol) and $PbI_2$ (50 mg/mL in dimethylformamide) were spin-coated onto the FTO-coated glass at 5000 RPM for 35 seconds. A post-spin anneal (50°C, 10 minutes) was performed on a hotplate.

*Electron spectroscopy measurements.* Partial electron yield (PEY) and total electron yield (TEY) nitrogen *K*-edge NEXAFS, PES, N *1s* RPES and RAES and normal N *KVV* AES were performed at the synchrotron BESSY II, at the soft X-ray beamline PM4 (operated in pseudo single bunch mode) and end-station LowDosePES [24]. The crystals were cleaved in UHV immediately prior to measurement. All measurements were performed at room temperature in a UHV chamber with a base pressure of $1\times10^{-9}$ mbar. Residual gas measurements indicate that water vapor is the largest contributor to the base pressure; carbon-based contaminants include carbon monoxide, carbon dioxide and methane. The cleanliness of the cleaved crystal surfaces was checked using PES with a photon energy of 535 eV. NEXAFS measurements were performed with a photon bandwidth/resolution of ~350 meV. TEY-NEXAFS was recorded by measuring the sample drain current and PEY-NEXAFS was simultaneously recorded by utilizing the Scienta Angle-Resolved Time-of-Flight (ARTOF) electron spectrometer as a simple electron-multiplier with a low kinetic energy cut-off of 50 eV. Photon flux normalization was performed with photo-current measurements recorded from the last beamline focusing mirror. The photon flux ranged between $10^7$ to $10^8$ photons sec$^{-1}$, distributed over a spot size of 100 x 300 µm$^2$, for the photon energies and beamline settings used in this work. Photon energy calibration was achieved by measuring the kinetic energies of Au *4f* photoelectrons (originating from a gold polycrystalline foil) excited by first and second order light and taking the difference, without changing the beamline or spectrometer settings. The NEXAFS background subtraction for MAPI was performed by first subtracting the reference N *K*-edge NEXAFS spectrum for PbI$_2$, which lacks nitrogen, followed by the subtraction of a linear background. Detailed PES measurements were performed with a photon energy of 535 eV, photon bandwidth of ~270 meV and an electron energy resolution of ~500 meV. Combined RPES, RAES and AES measurements were performed at several excitation energies around 400 eV, with a photon bandwidth of ~500 meV and an electron energy resolution of ~500 meV. The beamline monochromator was optimized to suppress higher-order excitation lines for the RAES measurements, to minimize the introduction of artifacts into the electron spectra. Binding energy calibration was performed by recording Au *4f* spectra from a reference gold foil and setting the binding energy of $E_{Au4f7/2}$ to 84.0 eV.

*Electronic structure and NEXAFS calculations.* AIMD simulations for supercells of MAPI were performed with the CP2K code [27]. NPT simulations at 300 K and 0 atm were performed for 50 ps with a 0.5 fs time-step using orthorhombic simulation cells with fixed cell side ratios matching those of the starting cell. Temperature was maintained using a 4-chain Nosé-Hoover thermostat with a coupling time of 20 fs. Forces for the dynamics were obtained using density functional theory (DFT) based electronic structure calculations, with the PBE exchange-correlation functional and Grimme's D3 van der Waals correction. The sampling of the electronic wave function was restricted to the Γ Brillouin zone point. The Gaussian and Plane Wave (GPW) method was used with a multi-grid consisting of 5 grids and a cutoff of 600 Rydberg in combination with Goedecker-Teter-Hutter (GTH) pseudopotentials. Corresponding basis sets used were TZVP-MOLOPT-GTH for C, N, and H, and DZVP-MOLOPT-SR-GTH for Pb and I. Initial cell parameters for MAPI were taken from Poglitsch and Weber's study on various perovskite compositions and phases [25]. A 2x2x2 supercell was generated from the unit cell of the tetragonal β phase (a = b = 8.855 Å, c = 12.659 Å). The MA$^+$ ions were centered in the cavities of the inorganic PbI$_3^-$ lattice with the N-C bond aligned along the c axis. To generate the computed NEXAFS spectrum and PDOS calculations, snapshots of the cell geometries were sampled for five configurations at regular intervals during the molecular dynamics simulations. Nitrogen *K*-edge NEXAFS spectra, obtained within the half core-hole transition potential approximation, were calculated for each nitrogen present in the supercell. Inner-shell spectroscopies are enabled by all-electron calculations in the framework of the Gaussian Augmented Plane Wave (GAPW) method and we used the same DFT functional and setting as in the GPW-based AIMD simulation with the following exceptions. All-electron basis sets used were 6-311++G2d2p for C, N, and H, whereas the TZVP-MOLOPT-SR-GTH pseudopotential basis sets were used for Pb and I. Gaussian convolution on the discrete transitions was performed. Each spectrum was then shifted by the energy of its core-excited N *1s* orbital to convert them into a consistent orbital energy frame.

The computed band-gap of MAPI (1.85 eV), as derived from the Kohn-Sham HOMO-LUMO energy difference, is over-estimated compared to the experimentally determined band-gap of 1.55 eV [6,44]. It is generally expected that the band-gaps obtained from DFT-based calculations with common approximate exchange-correlation functionals (i.e. LDA, GGA) are under-estimated due to the inability of DFT to reproduce the correct energy derivative discontinuity [45,46]. However, there is a fortuitous cancellation of errors due to GGA (PBE) under-estimation and the lack of spin-orbit coupling in our calculation, which leads to the over-estimation of the band-gap [47]. In addition, we note that the 1.85 eV band-gap is obtained for the finite-temperature MD trajectories and not for an optimized ground-state geometry. By convention, the binding energy of bound (i.e. occupied valence) states is defined as negative for electronic structure calculations, while for photoemission the binding energy is defined as positive. We use the PES convention throughout for consistency.

## X-ray diffraction measurements of MAPI single crystals

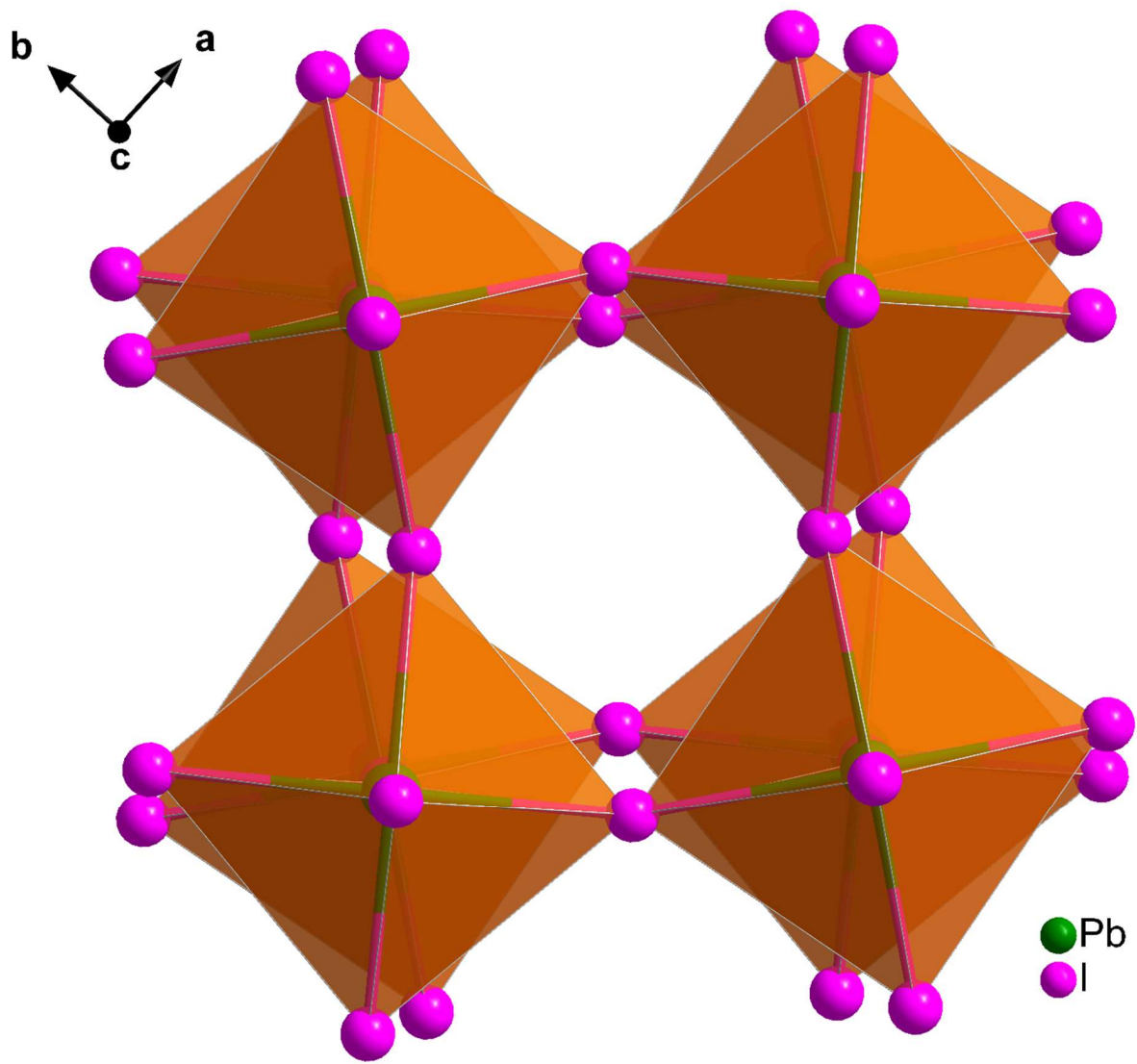

FIG S1. Refined crystal structure of the MAPI single crystal. The c-axis is pointing into the page. Carbon, hydrogen and nitrogen atoms have been omitted for clarity.

TABLE S1. Crystallographic parameter table.

| Space Group | a=b (Å) | c (Å) | α=β=γ (°) | Volume (Å$^3$) | Z | Reflections Collected | Independent reflections |
|---|---|---|---|---|---|---|---|
| $I4/m$ | 8.855(7) | 12.6590(15) | 90 | 992.61(17) | 4 | 2243 | 572 |

## Analysis of potential beam-induced and vacuum-induced chemical and electronic changes to the materials studied

Potential beam-induced damage to the sample(s) is monitored with Photoelectron Spectroscopy (PES). The typical markers of beam-induced chemical changes to HaP materials, as seen in PES measurements, include dose-dependent:

- loss of N *1s* photoelectron intensity relative to the Pb *4f* / *5d* / etc. photoelectron intensity,
- increases in the relative concentration of metallic lead,
- growth of peak asymmetry, reflecting the emergence of different chemical states, in one or more of the core level peaks,
- uncorrelated changes in the carbon, nitrogen, lead and iodine photoelectron intensities and
- uncorrelated shifts in the photoelectron peak positions [48].

Fig. S3 shows the PES spectra, recorded from the same spot of a cleaved MAPI single crystal surface before and after a > 2 hour NEXAFS measurement (also recorded from the same spot). Table S2 summarizes the curve-fitted parameters for the MAPI PES spectra displayed in Fig. S3 and Fig. S4(a) presents an additional core level spectrum (Pb *5d*). From the:

- absence of a metallic lead peak in the Pb *4f* spectra,
- similar before and after FWHM values (all core level spectra),
- similar values of $I_{After}$ / $I_{Before}$ (all core level spectra) and
- same ~0.2 eV shift in the peak positions of all core level spectra,

we conclude that all of the typical markers of beam-induced chemical changes are absent from the PES spectra recorded from MAPI, demonstrating that the chemical composition of the surface is intact after extended exposure to continuous X-ray irradiation and concomitant electron emission.

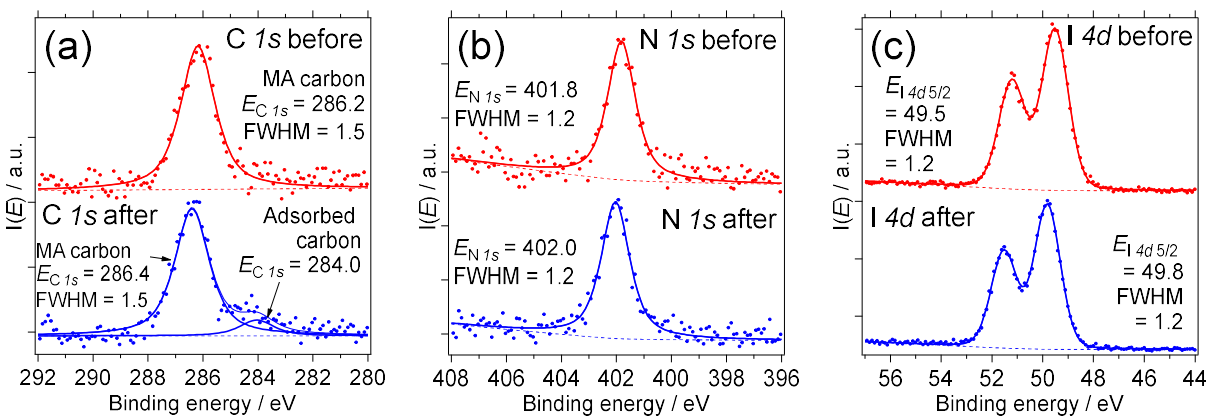

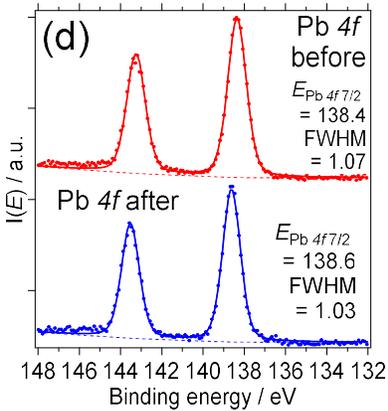

FIG. S2. (a)-(d) Photoelectron spectra of selected core levels representing all of the elements, except hydrogen, contained within the hybrid organic-inorganic lead halide perovskite. The spectra were recorded from the same spot, before and after a >2 hour NEXAFS measurement. Peak positions and full-width-half-maximum (FWHM) values are given in electronvolts.

During the >2 hour NEXAFS measurement, the C $1s$, N $1s$, I $4d$ and Pb $4f$ peak positions all show a ~0.2-0.3 eV shift towards higher binding energy. The same shift should be observed in the valence band spectra (Fig. S4(b)). However, the valence band spectrum of MAPI, recorded at hv = 535 eV, is noisy due to the spectrometer acquisition parameters used for this particular measurement. To accurately quantify the VBM position at hv = 535 eV, the $E_{Pb\ 5d5/2}$-$E_V$ offset is first derived from the valence band and shallow core spectra recorded at hv = 403.3 eV (Fig. S4(c,d)), which features a valence band spectrum recorded with better statistics. The application of this Kraut energy offset, which is material-dependent and independent of the photon/excitation energy, to the Pb $5d_{5/2}$ peak position, extracted from the Pb $5d$ spectrum recorded at hv = 535 eV (Fig. S4(a)), should yield the VBM positions for the hv = 535 eV spectra before and after the NEXAFS measurement [49]. The derived $E_{Pb\ 5d5/2}$-$E_V$ Kraut offset is (20.06 – 1.60) ~ 18.5 eV. The energetic position of the MAPI VBM was determined by approximating the low DOS of the VBM as an exponential function. This method has been shown to be more accurate than the conventional linear extrapolation method for this class of materials [6,16,50]. The Pb $5d_{5/2}$ peak positions for the spectra recorded at hv = 535 eV are 19.84 eV BE (before NEXAFS) and 20.06 eV (after NEXAFS). Hence, the VBM positions are (19.84 eV – 18.5) ~ 1.3 eV BE (before NEXAFS) and (20.06 – 18.5) ~ 1.6 eV BE (after NEXAFS), respectively. The MAPI VBM position shifts from ~1.3 eV below the Fermi level to ~1.6 eV below the Fermi level, indicating that the surface has been unintentionally n-doped. Given a reported transport band-gap of ~1.6 eV, we observe that the surface changes from moderately n-type to degenerately n-type with the Fermi energy positioned at the CBM [6].

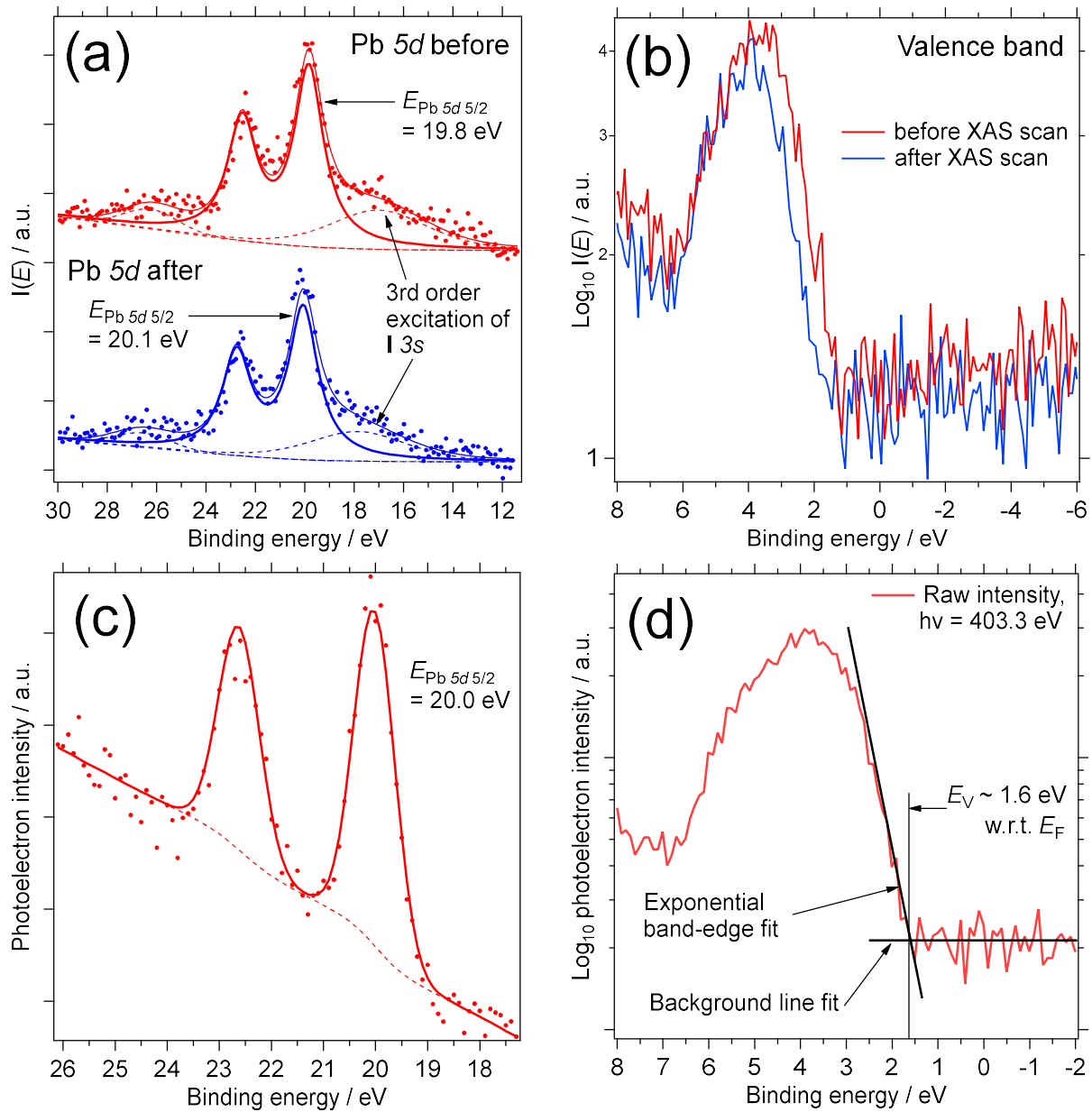

FIG. S3. Lead *5d* and valence band PES spectra of MAPI. (a,b) The Pb *5d* spectrum, utilized for chemical quantification, and the valence band spectrum are recorded at hv = 535 eV. The valence band spectra show a sloped background due to the measurement conditions used. **c,d** The Pb *5d* and valence band spectra are recorded at hv = 403.3 eV.

According to low photon flux UPS measurements (with minimal surface photovoltage effects) and Kelvin Probe measurements performed in the dark, MAPbX$_3$ surfaces are typically n-type, potentially indicating the presence of donor-type surface states [51]. The formation of donor-like metallic lead at levels undetectable with PES (e.g. part-per-million concentrations) may n-dope the perovskite. The decomposition pathway may first involve the decomposition of MAPI into MAI and PbI$_2$, followed by the loss of MAI and $I_2$ in vacuum. Contamination from the environment may also be responsible for the n-type surface. Physi-sorption of water in a $10^{-6}$ mbar vacuum has been correlated with the n-doping of MAPb($I_{3-x}Cl_x$) film surfaces [52]. While beam-induced chemical changes below the detection limit of our PES scans may have occurred, the adsorption of contaminants from the vacuum chamber at a pressure of ~$1\times10^{-9}$ mbar is a likely explanation for the surface n-doping. The C *1s* PES spectrum (Fig. S3(a)), recorded after the NEXAFS measurement, shows a small contribution from adventitious carbon; the carbon could originate from the adsorption of carbon monoxide/dioxide, methane, etc. Residual gas measurements indicate that water is the dominant contributor to the base pressure of the LowDosePES measurement chamber. The adsorption of water alone can explain the n-doping effect. From the standpoint of the measurements here, the ~0.2-0.3 eV shift of the Fermi energy is small and is not expected to affect the N *K*-edge NEXAFS, nor substantially affect the N *KVV* and N *1s* auto-ionization electron spectra. The electron spectra may be broadened if the Fermi energy position (and magnitude of the work function) changes gradually during the measurement. We conclude that only the doping level, and not the distribution of electronic states, of MAPI is altered by the experimental conditions since beam-induced chemical changes have been excluded and the same shift in the Fermi energy can be observed in the PES spectra of all core levels (Fig. S3).

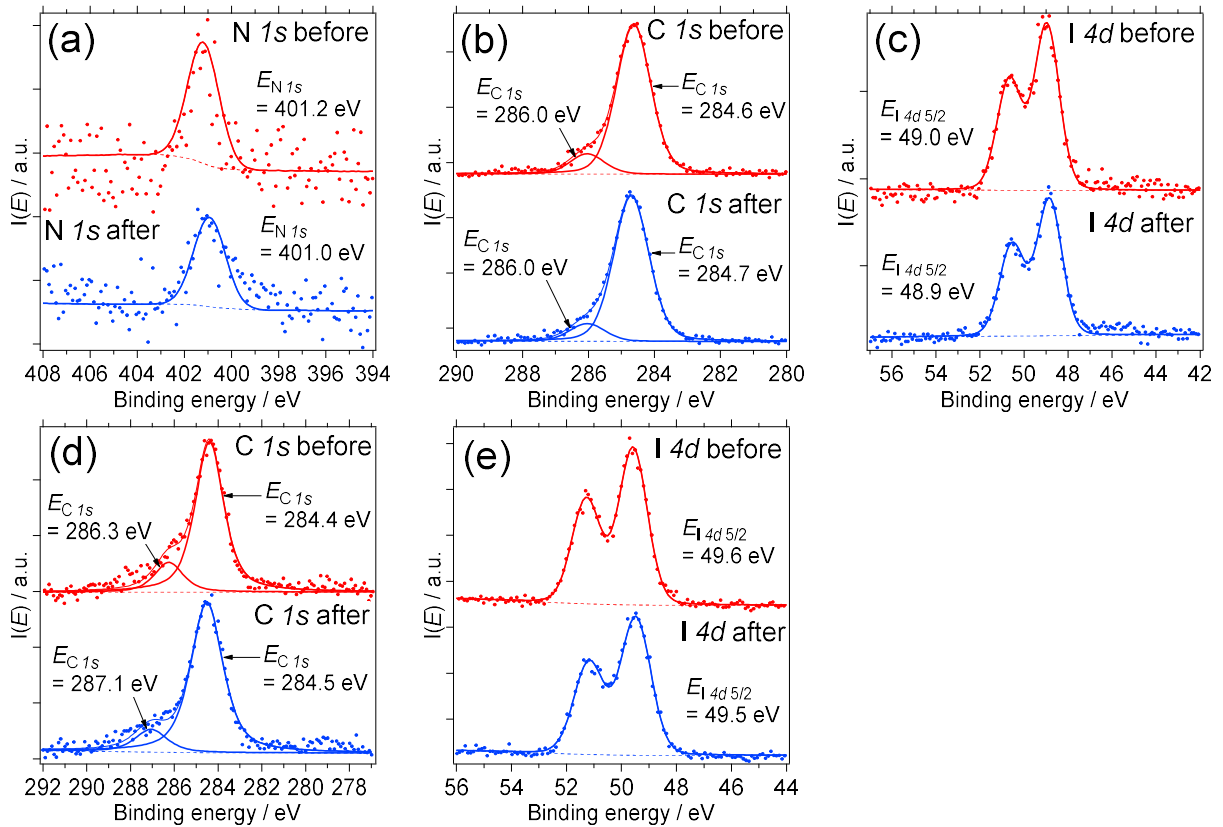

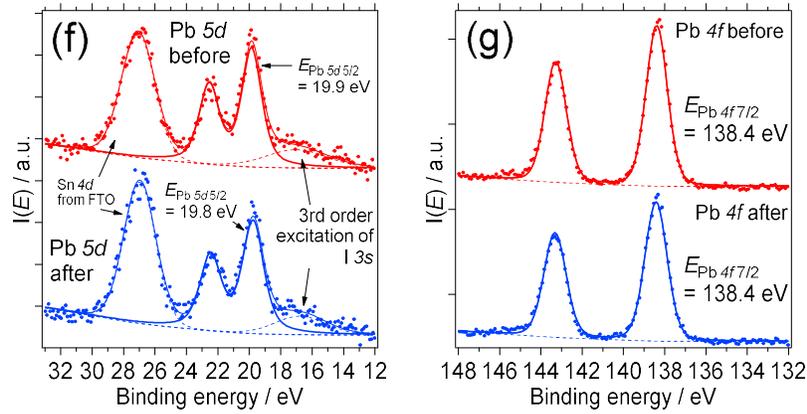

FIG. S4. Detailed PES spectra recorded from MAI and PbI$_2$. (a-c) MAI core level spectra. (d-g) PbI$_2$ core level spectra.

As an additional check for beam-induced decomposition of bulk MAPI into MAI and PbI$_2$, followed by the loss of MAI, we examine the direction of the shift of the Fermi energy. Given an inferred MAPI work function of ~3.6 eV and a reported work function of ~5.4 eV for PbI$_2$, this decomposition pathway should result in the Fermi energy moving away from the CBM due to the contact potential difference (CPD) [53]. The Fermi energy moves towards the CBM, which implies that the decomposition of MAPI into PbI$_2$ and MAI is unlikely.

An analysis of potential beam-induced changes, similar to the one described above for MAPI, in the binary precursor compounds MAI and PbI$_2$ leads us to the conclusion that the compounds are chemically stable under the X-ray beam. The supporting data, derived from curve-fits of the PES spectra displayed in Fig. S5, can be found in Table S2.

TABLE S2. Summary of parameters extracted from curve-fits of PES spectra recorded from MAPI, MAI and PbI$_2$. Adventitious is abbreviated as Adv. Before refers to PES measurements performed on the samples as soon as they have been introduced into the measurement chamber, while After refers to PES measurements performed at the same spot of the sample after >2 hours of continuous X-ray exposure (due to the NEXAFS measurement). Peak areas are labeled as I for intensity.

| Core level | | MAPI | MAI | PbI$_2$ |
|---|---|---|---|---|
| C 1s | Before | MA carbon<br>$E_{C1s}$ = 286.2 eV<br>FWHM = 1.5 eV<br>$I_{C1s}$ = 23.3 | Adv. carbon<br>$E_{C1s}$ = 284.6 eV<br>FWHM = 1.2 eV<br>$I_{C1s}$ = 77.6<br><br>Adv. + MA carbon<br>$E_{C1s}$ = 286.0 eV<br>FWHM = 1.2 eV<br>$I_{C1s}$ = 10.5 | Adv. carbon 1<br>$E_{C1s}$ = 284.4 eV<br>FWHM = 1.5 eV<br>$I_{C1s}$ = 56.5<br><br>Adv. carbon 2<br>$E_{C1s}$ = 286.3 eV<br>FWHM = 1.5 eV<br>$I_{C1s}$ = 11.2 |
| | After | Adv. carbon<br>$E_{C1s}$ = 284.0 eV<br>FWHM = 1.5 eV<br>$I_{C1s}$ = 2.6<br><br>MA carbon<br>$E_{C1s}$ = 286.4 eV<br>FWHM = 1.5 eV<br>$I_{C1s}$ = 20.7 | Adv. carbon<br>$E_{C1s}$ = 284.7 eV<br>FWHM = 1.3 eV<br>$I_{C1s}$ = 77.6<br><br>Adv. + MA carbon<br>$E_{C1s}$ = 286.0 eV<br>FWHM = 1.3 eV<br>$I_{C1s}$ = 9.4 | Adv. carbon 1<br>$E_{C1s}$ = 284.5 eV<br>FWHM = 1.7 eV<br>$I_{C1s}$ = 63.7<br><br>Adv. carbon 2<br>$E_{C1s}$ = 287.1 eV<br>FWHM = 1.7 eV<br>$I_{C1s}$ = 9.7 |
| | $I_{After} / I_{Before}$ | 0.9 (MA carbon only) | -- | -- |
| N 1s | Before | $E_{N1s}$ = 401.8 eV<br>FWHM = 1.2 eV<br>$I_{N1s}$ = 21.9 | $E_{N1s}$ = 401.2 eV<br>FWHM = 1.6 eV<br>$I_{N1s}$ = 6.4 | |
| | After | $E_{N1s}$ = 402.0 eV<br>FWHM = 1.2 eV<br>$I_{N1s}$ = 22.0 | $E_{N1s}$ = 401.0 eV<br>FWHM = 1.6 eV<br>$I_{N1s}$ = 4.7 | |
| | $I_{After} / I_{Before}$ | 1.0 | 0.7 | |
| Pb 5d | Before | $E_{Pb5d5/2}$ = 19.8 eV<br>FWHM = 1.4 eV<br>$I_{Pb5d5/2}$ = 11.4 | | $E_{Pb5d5/2}$ = 19.9 eV<br>FWHM = 1.5 eV<br>$I_{Pb5d5/2}$ = 11.3 |
| | After | $E_{Pb5d5/2}$ = 20.1 eV<br>FWHM = 1.4 eV<br>$I_{Pb5d5/2}$ = 9.7 | | $E_{Pb5d5/2}$ = 19.8 eV<br>FWHM = 1.6 eV<br>$I_{Pb5d5/2}$ = 11.4 |
| | $I_{After} / I_{Before}$ | 0.9 | | 1.0 |
| Pb 4f | Before | $E_{Pb4f7/2}$ = 138.4 eV<br>FWHM = 1.1 eV<br>$I_{Pb4f7/2}$ = 112.9 | | $E_{Pb4f7/2}$ = 138.4 eV<br>FWHM = 1.2 eV<br>$I_{Pb4f7/2}$ = 133.5 |
| | After | $E_{Pb4f7/2}$ = 138.6 eV<br>FWHM = 1.0 eV<br>$I_{Pb4f7/2}$ = 104.0 | | $E_{Pb4f7/2}$ = 138.4 eV<br>FWHM = 1.3 eV<br>$I_{Pb4f7/2}$ = 120.2 |
|  | $I_{After}$ / $I_{Before}$ | **0.9** |  | **0.9** |
| **I 4d** | Before | $E_{I4d5/2}$ = 49.5 eV<br>FWHM = 1.2 eV<br>$I_{I4d5/2}$ = 78.1 | $E_{I4d5/2}$ = 49.0 eV<br>FWHM = 1.3 eV<br>$I_{I4d5/2}$ = 14.0 | $E_{I4d5/2}$ = 49.6 eV<br>FWHM = 1.2 eV<br>$I_{I4d5/2}$ = 46.5 |
|  | After | $E_{I4d5/2}$ = 49.8 eV<br>FWHM = 1.2 eV<br>$I_{I4d5/2}$ = 68.5 | $E_{I4d5/2}$ = 48.9 eV<br>FWHM = 1.3 eV<br>$I_{I4d5/2}$ = 11.7 | $E_{I4d5/2}$ = 49.5 eV<br>FWHM = 1.3 eV<br>$I_{I4d5/2}$ = 43.0 |
|  | $I_{After}$ / $I_{Before}$ | **0.9** | **0.8** | **0.9** |

### Chemical composition and structure of the cleaved MAPI crystal surface

The chemical composition of the cleaved single crystal MAPI surface can be accurately quantified with experimentally calibrated sensitivity factors. We note that the as-cleaved crystal surface shows no observable adventitious carbon (Fig. S3(a)), which enables us to perform quantitative chemical analysis with higher accuracy. We estimate the I:Pb ratio for the MAPI surface by treating the binary precursor compound $PbI_2$ as a chemical standard, thus avoiding the need for accurate quantification of photon energy-dependent parameters such as the photoionization cross-sections, asymmetry parameters, transmission efficiency of the electron spectrometer, etc. Due to the low signal-to-background of the N *1s* PES spectrum of MAI (Fig. S5(a)), arising from adventitious carbon coating the film surface, that compound was not utilized as a chemical standard to derive the I:N ratio for MAPI. We first examine the PES survey scans of the cleaved single crystal MAPI and binary precursor thin film surfaces (Fig. S6) and find all of the expected elements. The $PbI_2$ sample is not expected to contain any metallic lead, which would oxidize to lead oxide upon air exposure, due to the 99.999% purity of the compound. However, the $PbI_2$ film is covered by adventitious carbon and its surface termination is unknown. Since beam-induced damage to $PbI_2$ and MAPI was found to be negligible, we arbitrarily take the curve-fitted parameters from the spectra recorded before the NEXAFS measurements to estimate the I:Pb ratio.

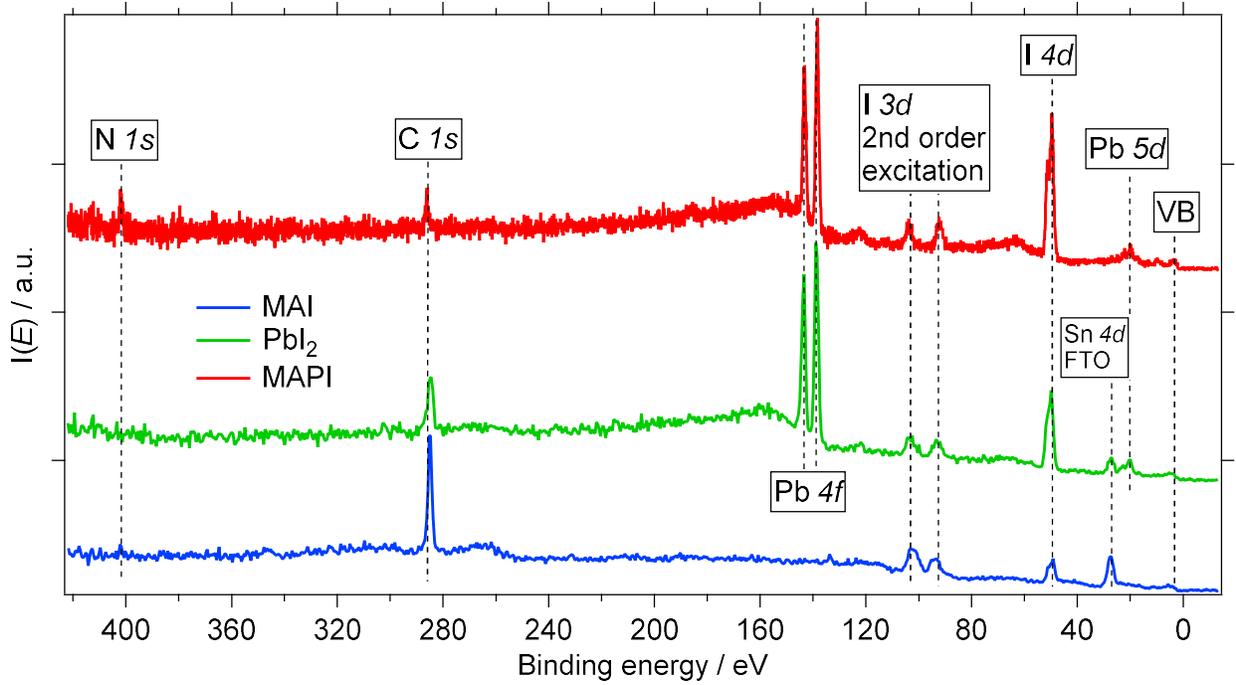

FIG. S5.  PES survey/overview spectra of MAI, PbI$_2$ and MAPI.  The spectra were recorded with a photon energy of 535 eV.  The PbI$_2$ and MAI films are ultra-thin, hence photoelectron peaks (e.g. Sn *4d*) from the underlying fluorine-doped tin oxide substrate are detected.

Starting with PbI$_2$, the curve-fitted Pb *5d$_{5/2}$* area is 11.3 and the I *4d$_{5/2}$* area is 46.5 (Table S2). Assuming a I:Pb = 2.0 ratio, the correction factor for I$_{Pb5d5/2}$ is 11.3 / ( 46.5 / 2 ) ~ 0.486.  The application of the derived correction factor to MAPI yields a I:Pb ratio of (78.1 / (11.4 / 0.486 ) ) ~ 3.3.  We use a structural model, with an exposed (001) cut (Fig. S7), to estimate the expected I:Pb ratio given the photon energy used (535 eV) and the resulting inelastic scattering length(s) of the Pb *5d* and I *4d* photoelectrons.  Using the software QUASES, which computes the IMFP based on the Tanuma-Powell-Penn (TPP-2M) formula, and PbI$_2$ as a proxy for MAPI, the IMFP of electrons photo-emitted from the shallow I *4d* and Pb *5d* core levels is ~ 14 Å  [54].  Tetragonal MAPI features a c-axis lattice parameter of ~12.7 Å.  Since ~95% of the photoelectron intensity originates from three IMFP lengths and the IMFP and c-axis lengths are comparable, the PES measurements are predominantly sensitive to the outermost three monolayers of bulk MAPI.  By taking one column of MAPI unit cells, aligned along the c-axis, assuming the surface is terminated by I$^-$ and MA$^+$ ions, and accounting for the exponential attenuation of the photoelectron signal, we count

$$N(iodide) = (1e^0) + \left(2e^{\frac{-6.4}{14.0}}\right) + \left(1e^{\frac{-12.7}{14.0}}\right) + \left(2e^{\frac{-1.1}{14.0}}\right) + \left(1e^{\frac{-25.4}{14.0}}\right) + \left(2e^{\frac{-3.8}{14.0}}\right) + \left(1e^{\frac{-38.1}{14.0}}\right) = 3.6$$

$$N(lead) = \left(1e^{\frac{-6.4}{14.0}}\right) + \left(1e^{\frac{-1.1}{14.0}}\right) + \left(1e^{\frac{-31.8}{14.0}}\right) = 0.99$$

atoms for an I:Pb ratio of 3.6. Using the same procedure, but assuming the surface is terminated by $Pb^{2+}$ and $I^-$ ions, we count

$$N(iodide) = (2e^0) + \left(1e^{\frac{-6.4}{14.0}}\right) + \left(2e^{\frac{-1.7}{14.0}}\right) + \left(1e^{\frac{-19.1}{14.0}}\right) + \left(2e^{\frac{-25.4}{14.0}}\right) + \left(1e^{\frac{-31.8}{14.0}}\right) + \left(2e^{\frac{-38.2}{14.0}}\right) = 4.3$$

$$N(lead) = (1e^0) + \left(1e^{\frac{-12.7}{14.0}}\right) + \left(1e^{\frac{-25.4}{14.0}}\right) + \left(1e^{\frac{-38.1}{14.0}}\right) = 1.6$$

atoms for an I:Pb ratio of 2.7. The experimentally derived I:Pb ratio of 3.3 is greater than the ratio of 2.7, hence we deduce that the surface is likely terminated by $I^-$ and $MA^+$, on average, over a length scale of hundreds of microns; this finding is consistent with calculations of the preferred thermodynamic surface [55]. A surface reconstruction and/or inhomogeneous cleaved single crystal surface may explain the ~10% deviation of the experimental I:Pb ratio (3.3) from the calculated ratio (3.6). We deduce that the surface region of the cleaved single crystal sampled by PES is chemically similar to the bulk.

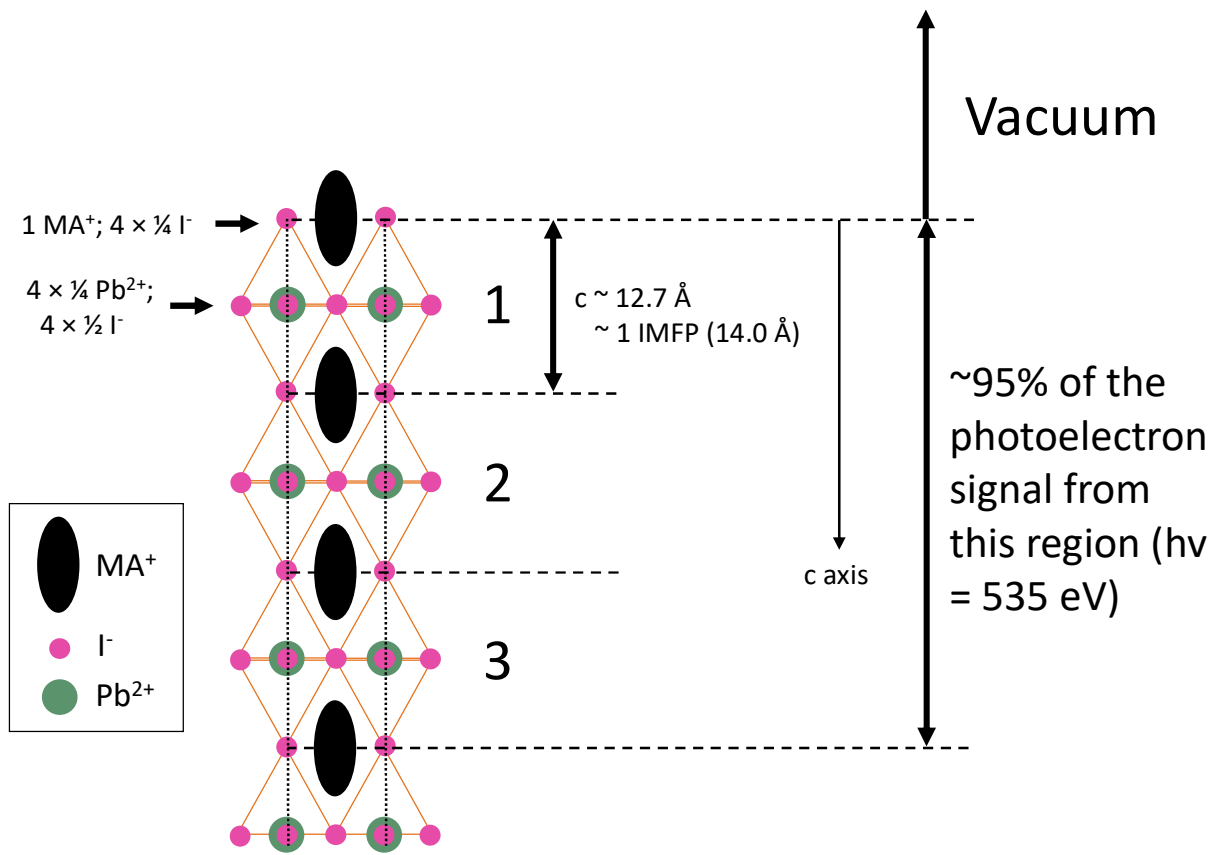

FIG. S6. Surface structural model of the MAPI crystal. One column of MAPI unit cells aligned along the c-axis and perpendicular to the surface is shown. A two-dimensional projection of the three-dimensional structure is shown for simplicity. The surface region probed with Pb *5d* and I *4d*

photoelectrons, generated by photo-absorption with hv = 535 eV, is essentially the same as the region spanned by three monolayers.

To confirm the chemical similarity of the surface with the bulk, we recorded two complementary NEXAFS measurements simultaneously: (a) more surface-sensitive partial electron yield (PEY) with a kinetic energy cut-off of 50 eV and (b) more bulk-sensitive Total Electron Yield (TEY).  Here we define the surface to be the depth from which information-containing Auger electrons are emitted.  Assuming the electrons Auger-emitted from methylammonium have comparable kinetic energies to the electrons Auger-emitted from aqueous ammonium (~340 to ~390 eV), we determine the sampling depth to be ~35 Å using similar reasoning as above [56].  The spatial depth sampled by Auger emission is comparable to the depth sampled by PES (with hv = 535 eV).  The as-recorded TEY-NEXAFS spectrum for MAPI is displayed in Fig. S8(a), along with the spectra for the binary precursors.  The Pb $N5$-edge ($4d_{5/2}$-to-$6/np*$) absorption threshold is close in energy to the nitrogen $K$-edge but does not overlap; it can be observed in the spectra of MAPI and PbI$_2$ but not of MAI.  We used the PbI$_2$ NEXAFS spectrum as a background level for the MAPI NEXAFS spectrum, and the subtraction of the PbI$_2$ NEXAFS spectrum removes the Pb $N5$-edge feature as shown in Fig. S8(b).  The PEY and TEY spectra shown in Fig. S8(b) overlap, indicating that the Auger and auto-ionization electron spectra presented in this work are representative of MA in MAPI.  Our N $K$-edge spectrum recorded from a cleaved MAPI single crystal surface is qualitatively similar to two published experimental spectra recorded from MAPI thin films [57,58].  In addition, the N $K$-edge NEXAFS spectrum of MAPI is qualitatively similar to N $K$-edge NEXAFS spectra recorded from ammonium in solvated environments, such as cationic glycine in an acidic aqueous solution and ethylammonium in ethanol, with some differences in the extent of the near-edge absorption features, likely due to differences in chemical bonding and in the electrostatic environments  [59,60].  This observation confirms that the nitrogen in MA is tetrahedrally-coordinated like ammonium and ethylammonium.

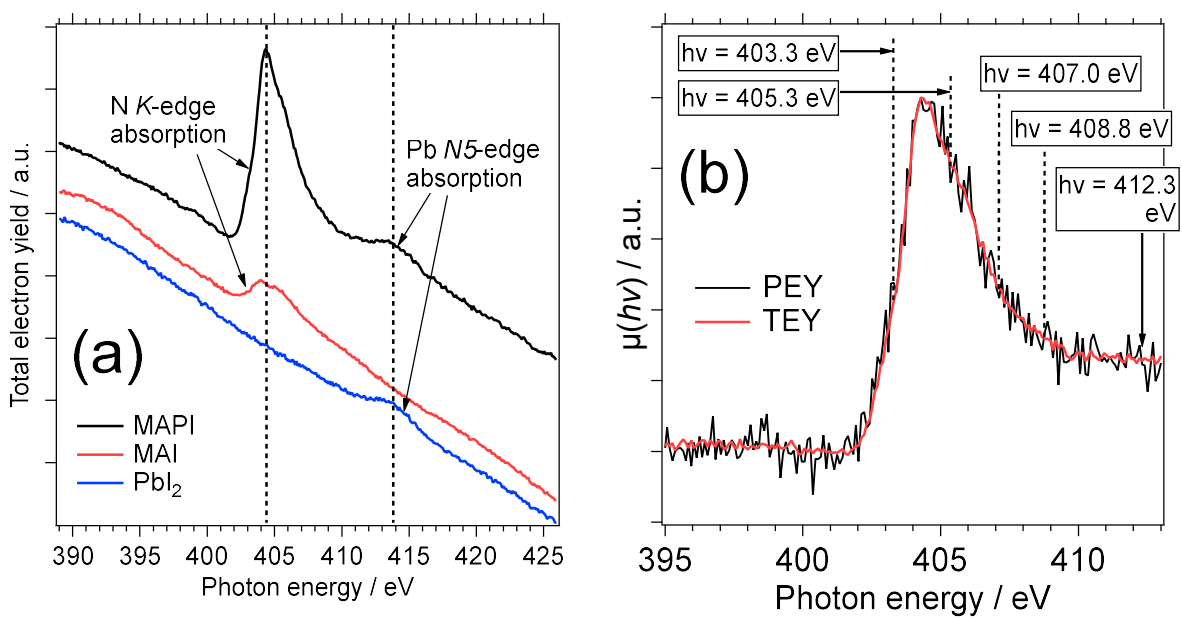

FIG. S7. Nitrogen *K*-edge NEXAFS spectra. (a) As-recorded TEY-NEXAFS spectra recorded from MAPI and its binary precursors. The background levels of the NEXAFS spectra have not been corrected yet. Pb 4d-to-6p* (*N5*-edge) absorption can be observed in the PbI$_2$ and MAPI spectra but not in the MAI spectrum. (b) Background-corrected PEY- and TEY-NEXAFS spectra recorded simultaneously from single crystal MAPI. Photon energies used to record detailed electron spectra arising from N *1s* core hole decay are labeled.

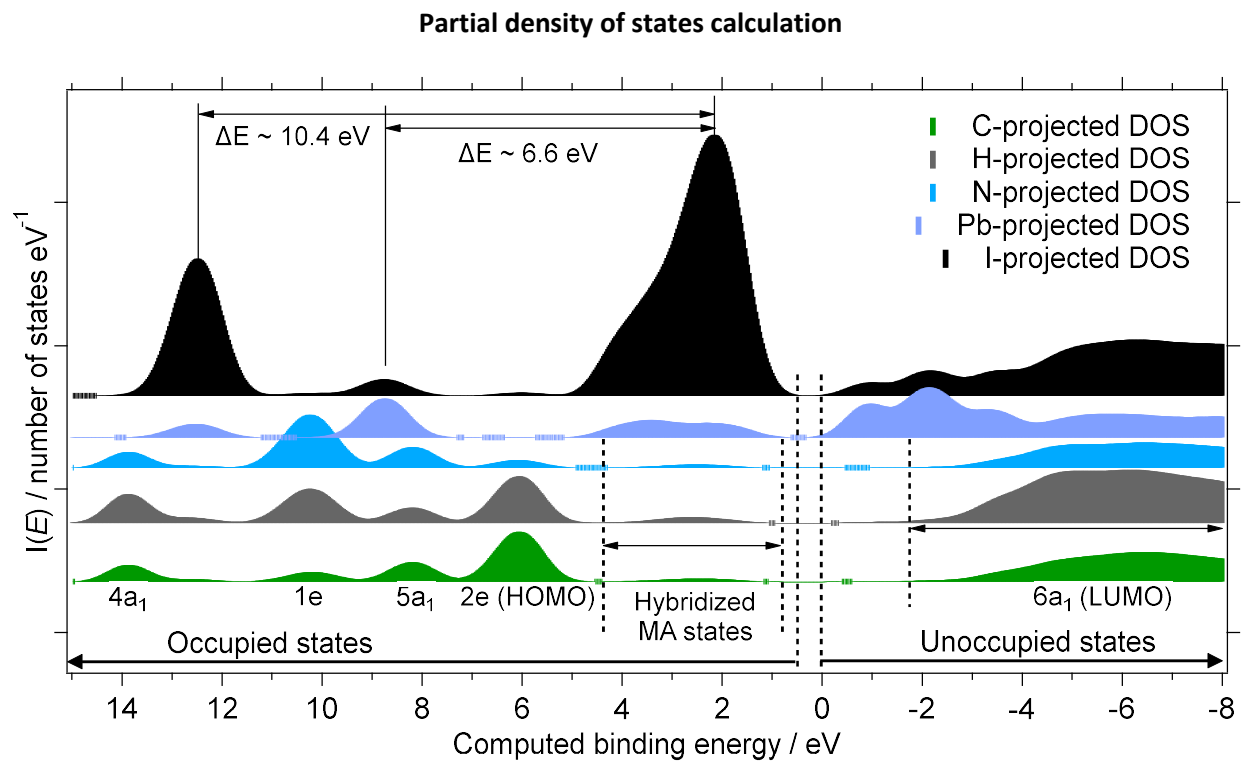

FIG. S8. Computed ground-state element-projected density of states.

### Extraction and identification of N *KVV* Auger and N *1s* auto-ionization features

A collection of electron spectra, recorded from the same spot of a cleaved single crystal MAPI surface at off-resonant and resonant excitation energies, is shown in Fig. S9(a) in the binding energy view. The removal of the normal photoelectron features, which highlights all electron emission associated with N *1s* core hole decay, is accomplished by treating the off-resonant spectrum (hv = 394.5 eV) as the background spectrum and subtracting it from the resonant spectra, as visualized in Fig. S9(b). The difference spectra are displayed in Fig. S9(d) in the kinetic energy view. Similar measurements on PbI$_2$ show no N *1s* core hole decay (Fig. S10), as expected.

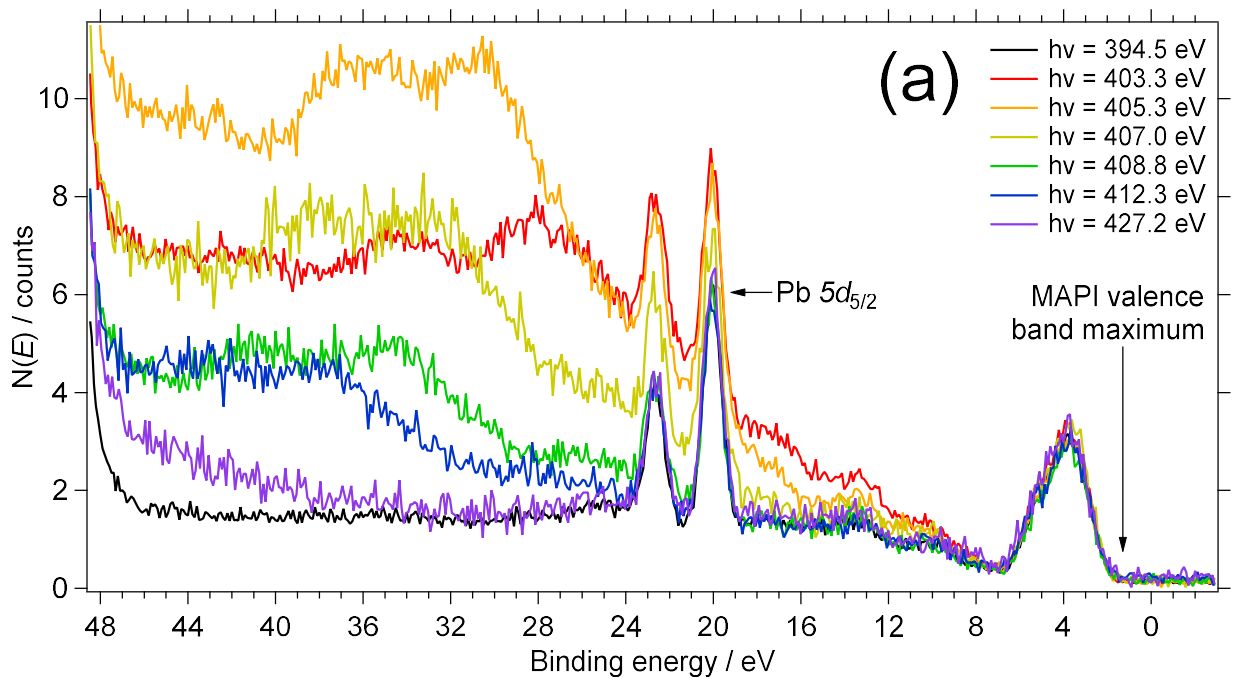
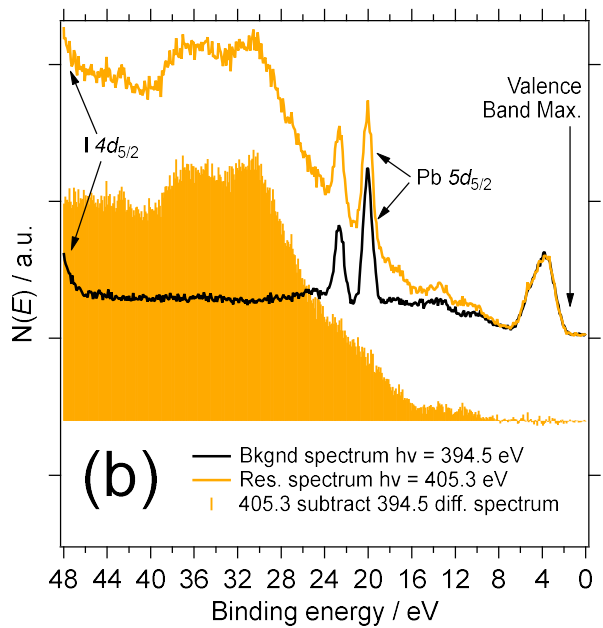
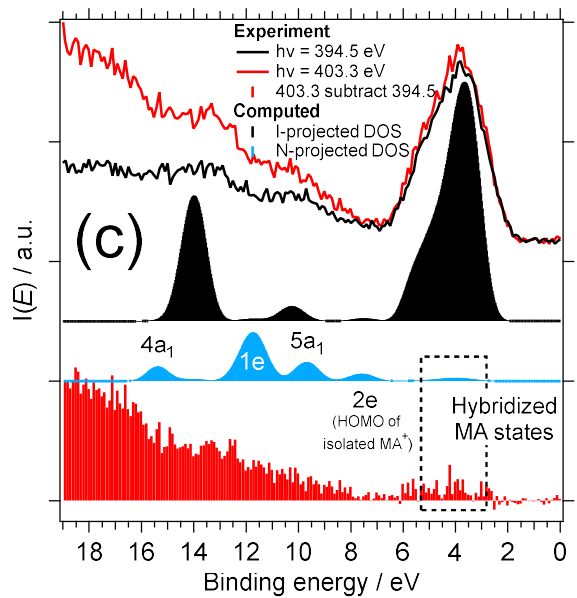

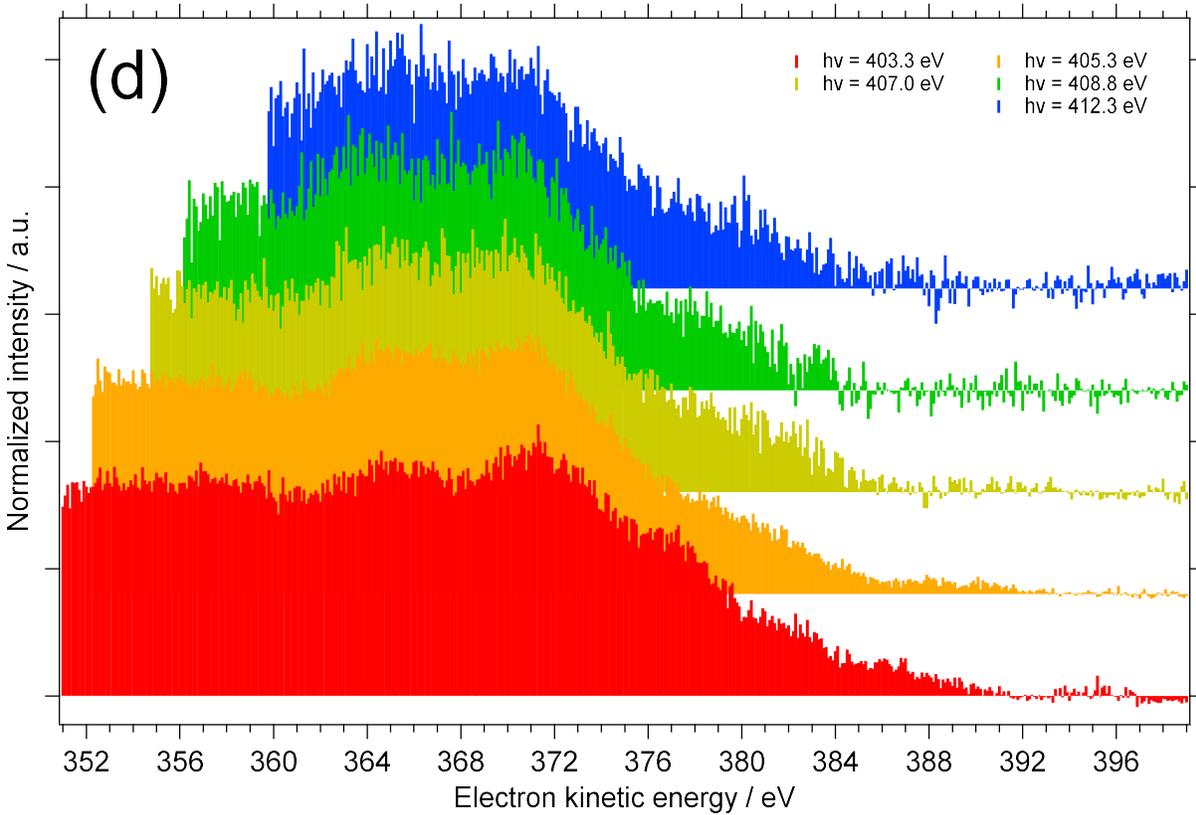

FIG. S9. Collection of combined photoelectron, Auger and auto-ionization electron spectra recorded from MAPI. (a) The spectra, measured as a function of electron kinetic energy, were normalized to the photon flux and aligned onto the binding energy scale using Fermi energy calibration. The off-resonant spectra were recorded at hv = 394.5 eV, hv = 412.3 and 427.2 eV. (b) The normal/non-resonant photoelectron features (e.g. Pb *5d*) are removed via the subtraction of the off-resonant hv = 394.5 eV spectrum, yielding a difference spectrum which highlights all electron emission associated with N *1s* core hole decay. (c) The difference spectrum extracted from the hv = 403.3 eV resonant spectrum is compared to the computed nitrogen-projected DOS. (d) Collection of N *1s* Auger and auto-ionization electron spectra, aligned onto the kinetic energy scale.

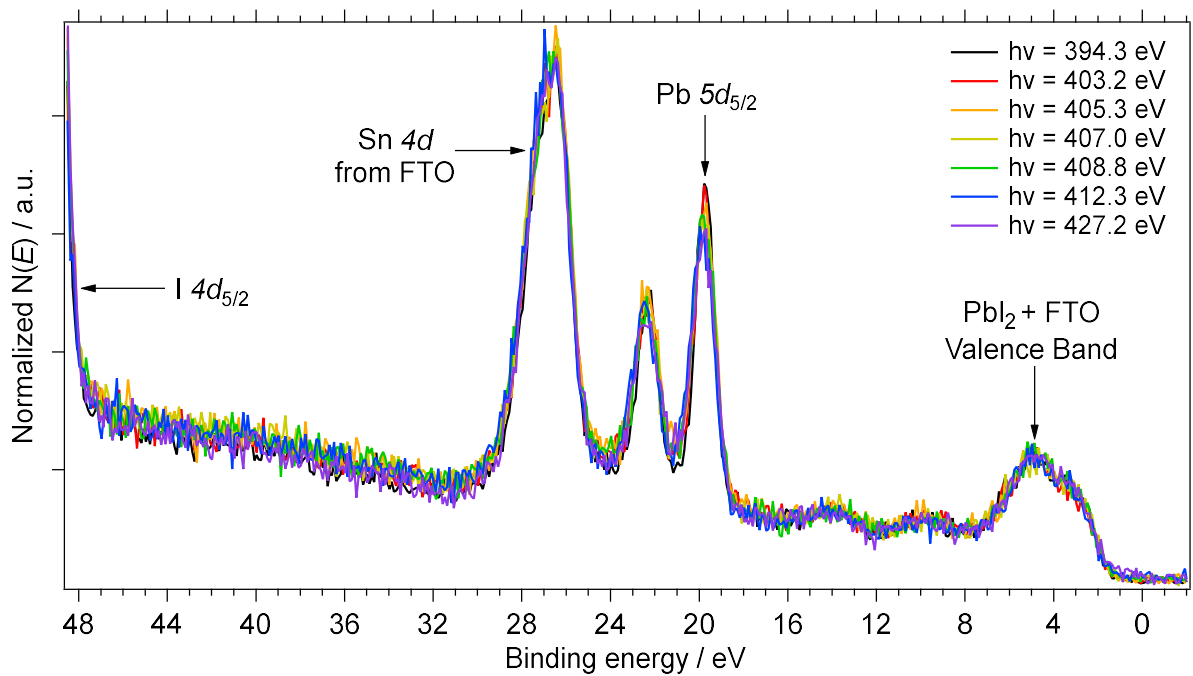

FIG. S10. Collection of electron spectra recorded from PbI$_2$ using the same photon energies used to record the off-resonant and resonant spectra from MAPI. The PbI$_2$, which does not contain nitrogen, shows no enhancement on-resonance.

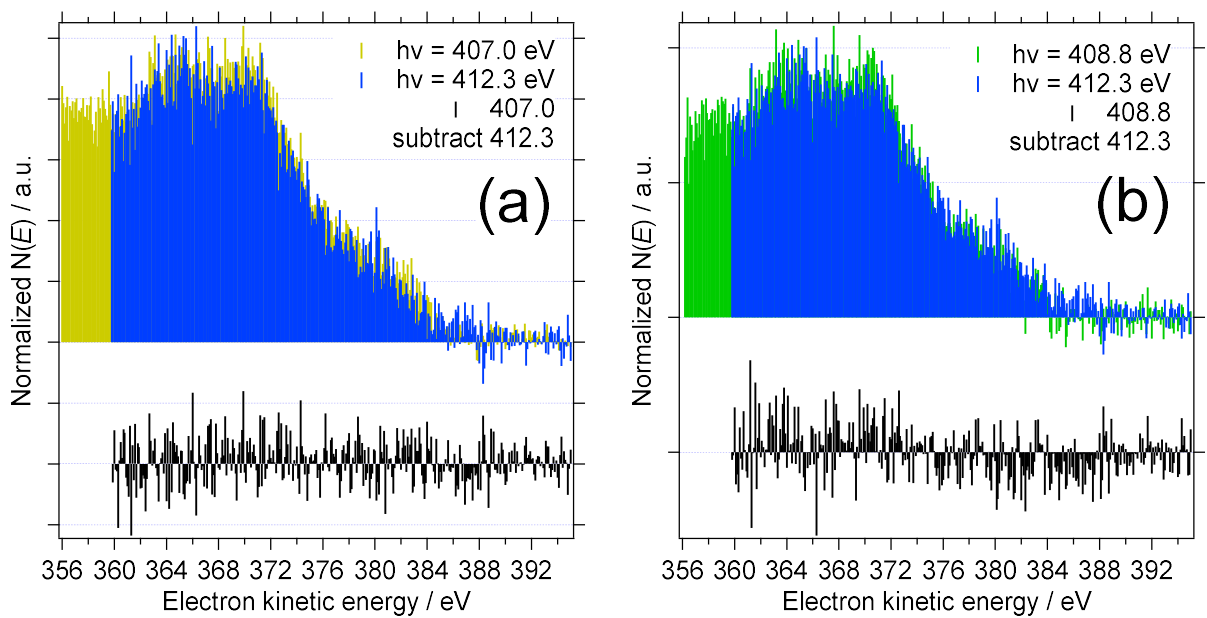

FIG. S11. Difference spectra. a,b Subtraction of an off-resonant Auger spectrum (hv = 412.3 eV) from intensity-scaled resonant Auger spectra (hv = 407.0, 408.8 eV).

# Nitrogen 1s core hole decay arising from resonant excitations to bound states

At least three different nitrogen 1s core hole decay pathways exist, for resonant excitations to bound states. These are shown in Fig. S12.

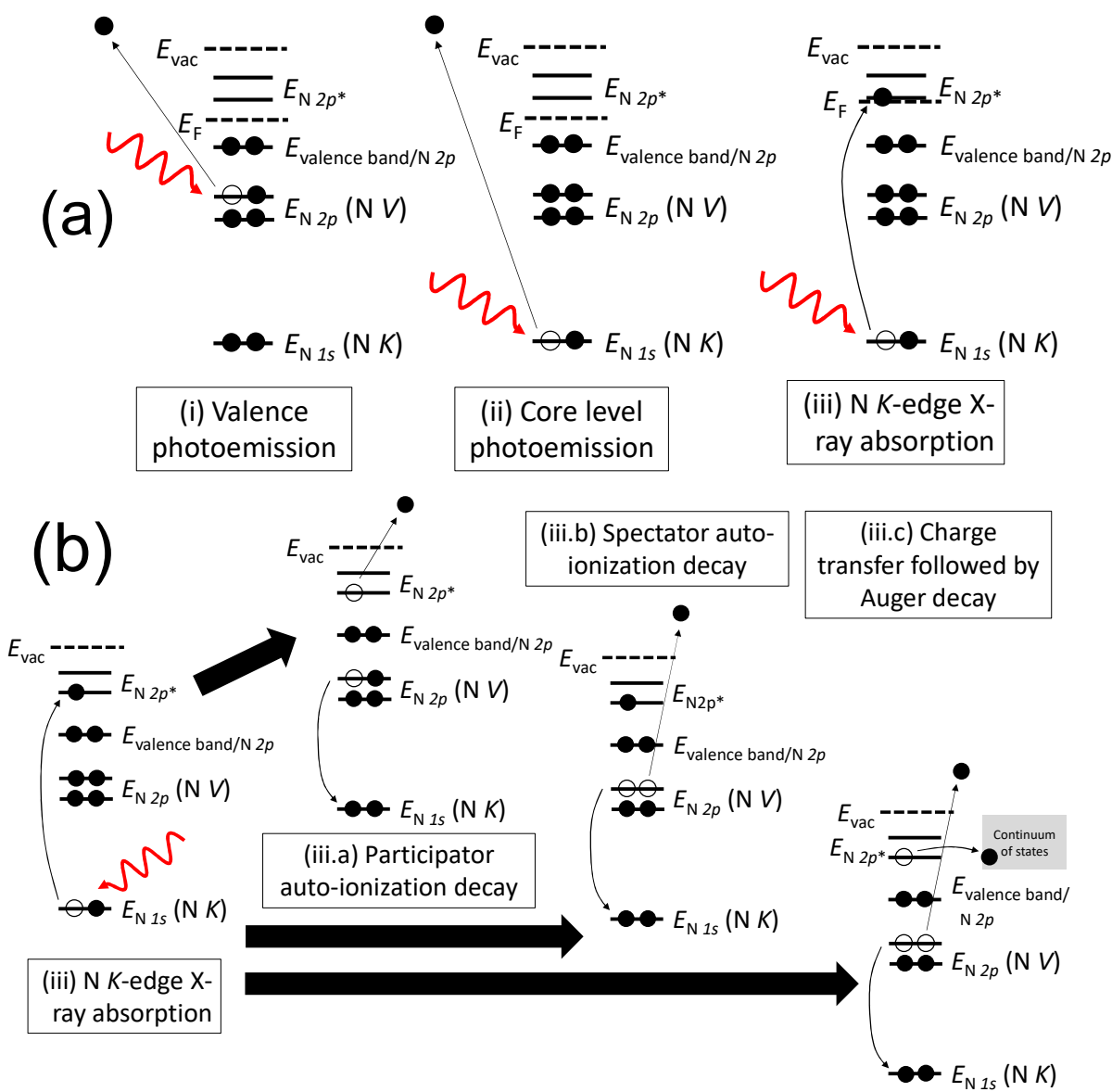

FIG S12. X-ray induced electron emission processes, in the one-electron picture, analyzed in this work. (a) Upon photo-absorption, transitions to the (i, ii) continuum or a (iii) bound state can occur, depending on the magnitude of the photon energy. (b) Upon the generation of a N 1s core hole following a transition to a bound state, specific decay processes will occur which may (iii.a, iii.c) involve the resonantly excited electron or not (iii.b).

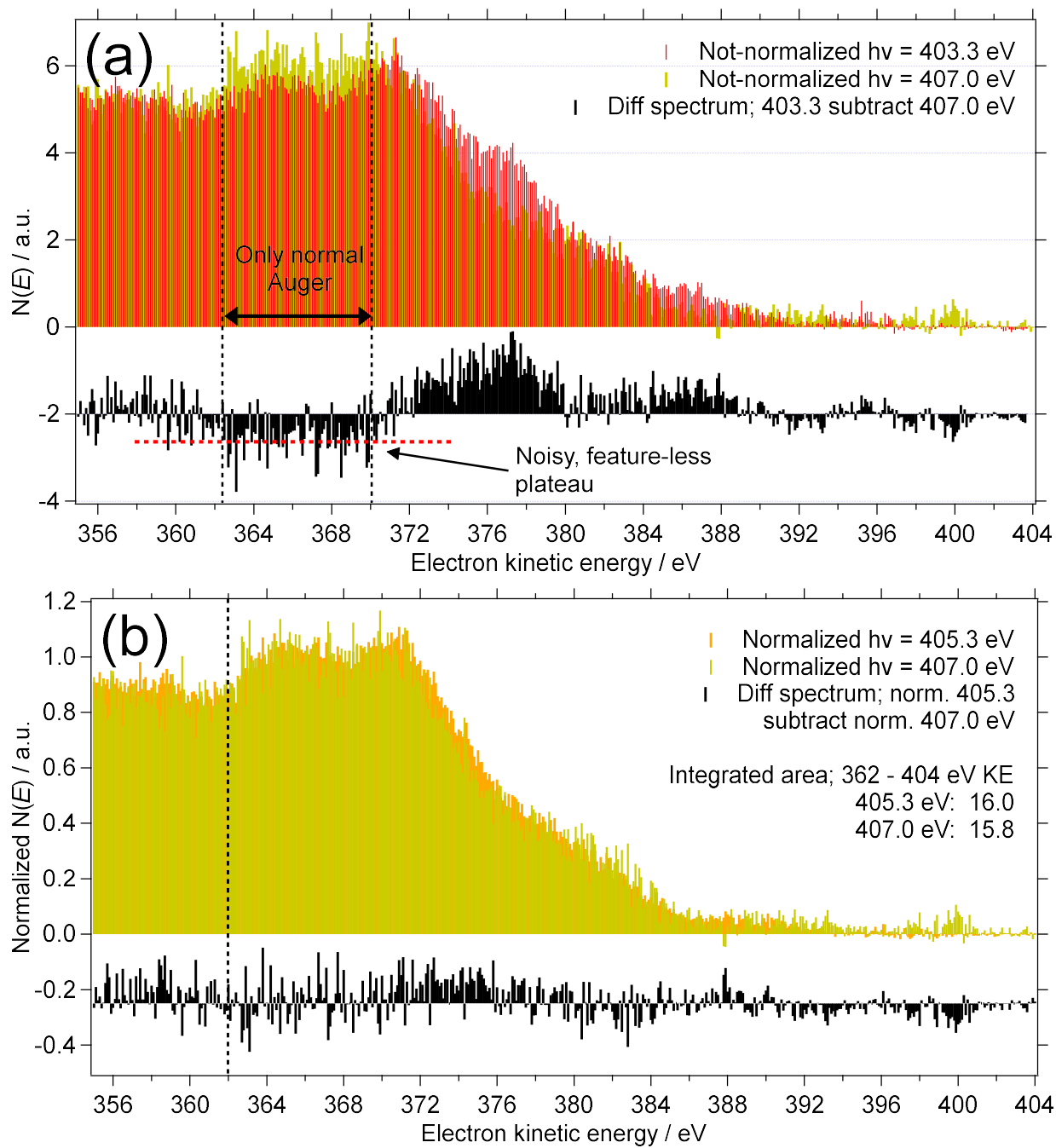

Fig. S13. Nitrogen *1s* core hole decay spectra associated with excitations to bound states. The spectrum generated by excitations to continuum states (hv = 407.0 eV) is treated as the normal Auger spectrum (see text). The difference spectrum highlights both the participator and spectator auto-ionization features. The integrated areas of normal Auger decay and auto-ionization decay, along with the nitrogen 1s core hole lifetime, are used to estimate the charge transfer time.